\documentclass[twocolumn,twoside,slac_two]{revtex4}
\usepackage{natbib}
\usepackage{graphicx}
\usepackage{epsfig}
\usepackage{fancyhdr}
\begin{document}

\title{$CP$ violating anomalous top-quark couplings at the LHC}

\author{Sudhir Kumar Gupta}
%\footnote{Speaker}, Alaettin Serhan Mete and G. Valencia}

\email{skgupta@iastate.edu}
%, serhan@iastate.edu, valencia@iastate.edu}
\affiliation{Department of Physics, Iowa State University, Ames, IA 50011.}

\begin{abstract}
In this talk, I review the $T$ odd correlations  induced by $CP$ violating anomalous top-quark couplings at both production and decay level in the process $gg \to t\bar{t} \to (b\mu^+ \nu_\mu) (\bar{b}\mu^- \bar{\nu}_\mu)$. In addition I will also focus on experimental sensitivities corresponding to the anomalous  couplings at the LHC. 
\end{abstract}

%\pacs{PACS numbers: 12.15.Ji, 12.15.Mm, 12.60.Cn, 13.20.Eb,13.20.He, 14.70.Pw}

\maketitle

\section{Introduction}

The standard Model of particle physics have been a successful theory to describe the observed experimental data till date. However there are theoretical issues such as matter-antimatter 
asymmetry and experimental issues such as finiteness of electric dipole moment (EDM) and anomalous magnetic moment which hints  for beyond standard model physics in the form of CP violation.
Though small in nature, as hinted by the present experimental bounds, it can be probed  by  proper choice of momentum dependent  CP violating observables.

Because of small nature of CP violating effects, it it worthwhile to consider processes 
which could be helpful in measuring these effects directly. One such possibility to test such direct effects is the study of processes involving tops. This is because of (a) top-pair events are highly abundant at the LHC with a rate of  $\sim 10^7$ pairs per year,  and also,  (b) top are short lived with a life-time smaller by one order of the QCD scale and hence they decay much before hadronization.

\section{CP Violation in top sector}
In a model independent approach without having any additional non-standard particle contents, the $CP$ violation is parametrized by anomalous 
top-quark couplings affecting both the production and decay vertices \cite{german}.  The $t\bar{t}$ production process is modified relative to the SM by the interaction

\begin{eqnarray}
{\cal L}_{cdm}&=&-ig_s\frac{\tilde{d}}{2}\bar{t}\, \sigma_{\mu\nu}\gamma_5  \, G^{\mu\nu}\, t,
\label{dtilde}
\end{eqnarray}
where $g_s$ is the strong coupling constant and $G^{\mu\nu}$ is the usual gluon field strength tensor. 
For the decay process $t\to b W^+$ , the most general decay vertex takes the following form~\cite{Antipin:2008zx}

\begin{equation}
\begin{array}{rcl}
%\begin{eqnarray}
\Gamma^\mu_{Wtb} &=& 
-\frac{g}{\sqrt{2}} \, V_{tb}^\star \,\bar{u}(p_b) [ \gamma_\mu (f_1^L P_L+f_1^R P_R) \\
&&-i  \sigma^{\mu\nu} (p_t-p_b)_\nu (f_2^L P_L+f_2^R P_R) ]  u(p_t). 
\label{ftilde}
%\end{eqnarray}
\end{array}
\end{equation}

We use $V_{tb}\equiv 1$, $f_1^L=1$, $f_1^R=0$ and $f_2^L=0$ as in the SM, and $f_2^R=f\exp{i(\phi_f+\delta_f)}$,  with  $\phi_f$ and $\delta_f$ as  $CP$ violating and CP conserving phases respectively. In a similar way $\bar{t}\to \bar{b} W^-$ decay vertex can also be written.

At the LHC, most of the top pair production events are due to gluon-fusion, which amounts to $\sim 90 \%$ of the total top-pairs. Therefore throughout this talk,  we consider only on the subprocess involving gluons as initial state.  Also since we found that final dimuonic mode is the cleanest at the LHC,  we restricted our studies to the process: $gg \to t\bar{t} \to (b\mu^+ \nu_\mu) (\bar{b}\mu^- \bar{\nu}_\mu)$.

For the aforementioned process, the relevant correlation which does not require intermediate state reconstructions, as obtained in ref~\cite{ours} are

\begin{eqnarray}
\tilde {\cal{O}}_1 &=& \epsilon(p_b,p_{\bar{b}},p_{\mu^+},p_{\mu^-}) \nonumber \\
\tilde {\cal{O}}_2 &=& \, \tilde{q}\cdot (p_{\mu^+}-p_{\mu^-}) \,\epsilon(p_{\mu^+},p_{\mu^-},p_b+p_{\bar{b}},\tilde{q}) \nonumber \\
\tilde {\cal{O}}_3 &=& \, \tilde{q}\cdot (p_{\mu^+}-p_{\mu^-}) \,\epsilon(p_{b},p_{\bar b},p_{\mu^+}+p_{\mu^-},\tilde{q}),
\label{prodcoprime}
\end{eqnarray}

The integrated asymmetries $A_i$,  corresponding to observables $O_i$
integrated counting asymmetries, $A_i$, can be defined as

\begin{eqnarray}
A_i &\equiv & \frac{N_{events}({\cal O}_i >0)-N_{events}({\cal O}_i<0)}{N_{events}({\cal O}_i>0)+N_{events}({\cal O}_i<0)}.
\label{asym}
\end{eqnarray}

It is Interesting to note that, from the experimental perspective, $\tilde {\cal{O}}_2 $ is most desirable as it does not require distinction between b and anti-b jets.
%--------------------------------------------------------------------   
\begin{figure*}
\centering
\resizebox{!}{7cm}{\includegraphics{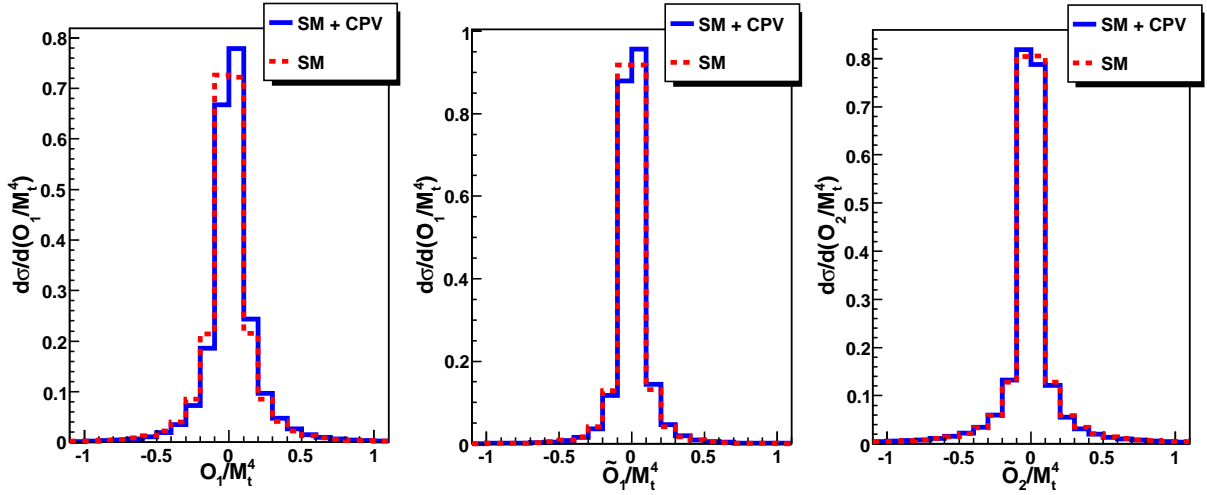}}
\caption{Differential distributions corresponding to observables ${\cal O}_1$ for $\tilde{d}=5\times  
10^{-4}~{\rm GeV}^{-1}$, ${\cal O}_{1,2}$ for $f\sin\phi_f= 5 \times 10^{-4}~{\rm GeV}^{-1}$  (solid lines) and the standard model (dotted line). Figures are taken from the Ref~\cite{ours}.}
\label{f:fig1}
\end{figure*}
%--------------------------------------------------------------------

When CP-violation occurs due to top decays, the spin and color averaged matrix element squared containing the $T$-odd correlations can be written as~\cite{Antipin:2008zx}

\begin{eqnarray}
|{\cal M}|^2_{T} &=& \, 
f\sin(\phi_f+\delta_f)\, \epsilon(p_t,p_{ b},p_{\ell^+},Q_{t})  \nonumber \\
&&
+  f\sin(\phi_f-\delta_f) \,\epsilon(p_{\bar t},p_{ \bar{b}},p_{\ell^-},Q_{\bar{t}}) .
 \label{asymcpdec}
\end{eqnarray} 

In this expression, first term contains information due to one decay vertex while the other term due to other. Clearly, we can not compare top and anti-top decays with the use of observables mentioned earlier in this talk. 
This means we need to define additional observables which could measure asymmetries corresponding to CP violating phases as well as CP conserving phases. Observables corresponding to CP violating phases are~\cite{ours} 

\begin{eqnarray}
{\cal O}_4 &=& \epsilon(P,p_b-p_{\bar{b}},p_{\mu^+},p_{\mu^-}) \nonumber \\
{\cal O}_5 &=& \epsilon(p_t,p_{\bar t},p_b+p_{\bar{b}},p_{\mu^+}-p_{\mu^-}) \nonumber \\
{\cal O}_6 &=&\,(t-u) \,
\epsilon(P,p_b+p_{\bar{b}},p_{\mu^+}-p_{\mu^-},q).
\label{decayco}
\end{eqnarray}

In the later part of this talk I will mention that we need to define a couple of more observables which carry information on strong CP violating phases.  

\section{Numerical Analysis}

We studied CP violating asymmetries using MadGraph~\cite{Alwall:2007st}. Our strategy was as follows: We first generated SM matrix-element squared for the aforementioned process and then add the contributions 
due to CP violating part as mentioned in Ref.~/cite{german}. 

We noted that the major background for the process under consideration is due to $pp \to b\bar{b} \mu^+ \mu^- X$,  with minimal acceptance cuts the cross-sections at the LHC are 4.3 pb and 24 pb which after implementing the basic cuts mentioned in the Ref.~/cite{ours} and a missing energy cut  $E_T{\!\!\!\!\!\!/\!\!\ }~ > 30 ~{\rm GeV}$ 
reduce to 2.3 pb and 0 respectively.

In order to estimate asymmetries, we generated $10^6$ events for each of the four cases: $\tilde{d} = 5 \times 10^{-4}~{\rm GeV}^{-1}$; $f\sin\phi_f= 5 \times 10^{-4}~{\rm GeV}^{-1}$; $f\sin\delta_f = 5 \times 10^{-4}~{\rm GeV}^{-1}$ and $\tilde{d} = f =0$, which correspond to  $CP$ violation in the production vertex, $CP$ violation in the decay vertex, strong phases in the decay vertex and the lowest order SM respectively.

The largest asymmetry for a $\tilde{d}$ coupling is $\tilde{A}_1$ but this is helpful only if we could distinguish clearly between b and anti-b jets. 
In  Figure~\ref{f:fig1}  we show the differential distributions
$d\sigma/d{{\cal O}}_1$ and $d\sigma/d{\tilde{\cal O}}_1$ for
$\tilde{d}= 5 \times 10^{-4}~{\rm GeV}^{-1}$ as well as
$d\sigma/d{\tilde{\cal O}}_2$ for $f\sin\phi_f= 5 \times 10^{-4}~{\rm
GeV}^{-1}$.

\subsection{LHC sensitivity}

Using the asymmetries as estimated in \cite{ours}, we calculated LHC sensitivities corresponding to the 
dimensionless anomalous couplings
\begin{eqnarray}
d_t \equiv  \tilde{d} \, m_t, && 
f_t \equiv  f\, m_t
\label{dimen}
\end{eqnarray}

We found that with one year of LHC run (which corresponds to integrated lumionosity = $10~fb^{-1}$), 5$\sigma$ sensitivity requires $A_i \geq 0.033$. Using this, we obtain
\begin{eqnarray}
|d_t| \geq  0.05,    && |\tilde{d}| \geq    3.0 \times 10^{-4}~{\rm GeV}^{-1}
  \\ \nonumber
|f_t \sin\phi_f| \geq   0.10,  && |f\sin\phi_f| \geq  6.0 \times 10^{-4}~{\rm GeV}^{-1}
\end{eqnarray}

It is to be noted that we have already checked our results for top anomalous couplings at production level against literature~\cite{sjolin} and found it consistent.

\subsection{Strong Interaction Phases}
In order to isolate true $CP$ violation from the $T$ -odd triple products, we construct two CP-even observables which is based on the most sensitive observable to phases in decay vertex. These are

\begin{eqnarray}
{\cal O}_a &=& \, \tilde{q}\cdot (p_{\mu^+}+p_{\mu^-}) \,\epsilon(p_{\mu^+},p_{\mu^-},p_b+p_{\bar{b}},\tilde{q}) \nonumber \\
{\cal{O}}_b &=& \, \tilde{q}\cdot (p_{\mu^+}-p_{\mu^-}) \,\epsilon(p_{\mu^+},p_{\mu^-},p_b-p_{\bar{b}},\tilde{q}).
\label{cpeven}
\end{eqnarray}

Proceeding as in the previous section we found that the larger of these two asymmetries we can write
\begin{eqnarray}
A_b &=& -0.32 f_t \sin\delta_f,
\label{cpeveneq}
\end{eqnarray}
from which we conclude that the LHC with $10~{\rm fb}^{-1}$ will have a $5\sigma$ sensitivity to  
\begin{eqnarray}
|f_t\sin\delta_f| \geq  0.10  && |f\sin\delta_f| \geq 6.0 \times 10^{-4}~{\rm GeV}^{-1}.
\end{eqnarray}

\section{Summary and Conclusion}

In this talk, we discussed the following issues:
\itemize
\item
We have estimated asymmetries due to anomalous top quark coupling both production and decay level.
\item
LHC sensitivities corresponding to these couplings has also been estimated using the asymmetries.
\item 
We also noted that the true CP phases can be isolated from the strong interacting phases.

\begin{acknowledgments}
This talk is based on a work done with A. S. Mete and G. Valencia~\cite{ours}. The work was supported in part by DOE under contract number DE-FG02-01ER41155. 
 
\end{acknowledgments}

\end{document}